\documentstyle[preprint,eqsecnum,aps]{revtex}
\begin{document}
\preprint{DOE/ER/40561-235-INT95-00-108}
\title{Low-energy QCD: Chiral coefficients\\
and the quark-quark interaction}
\vspace{0.5in}
\author{M.R. Frank$^\ast $ and T. Meissner$^\dagger $}
\address{
$\ast $Institute for Nuclear Theory, University of Washington, Seattle,
WA 98195}
\address{$\dagger $Department of Physics and Astronomy, University of
South Carolina, Columbia, SC 29208}
\maketitle
\vspace{0.5in}
\begin{abstract}
A detailed investigation of the low-energy chiral
expansion is presented within
a model truncation of QCD.  The truncation allows for a
phenomenological description of the quark-quark interaction in a
framework which maintains the global symmetries of QCD and permits a
$1/N_c$ expansion.
The model dependence of the chiral coefficients
is tested for several forms of the quark-quark interaction by varying
the form of the running coupling, $\alpha (q^2)$, in the infrared region.
The pattern
in the coefficients that arises at tree level is consistent with
large $N_c$ QCD, and is related to the model truncation.
\end{abstract}

\section{Introduction and Summary}

Phenomenological approaches to quantum chromodynamics(QCD)
continue to provide useful intuition into
the nature of the strong interaction, and compliment the more direct
evaluation via lattice techniques.  The utility of these
treatments is perhaps most
apparent in the study of chiral observables where
lattice calculations are subject to large uncertainties due to the
extrapolation to light quarks.  At low energies, this aspect of QCD is
characterized by chiral perturbation theory($\chi $PT)
\cite{chipt,ecker,ugmeissner}.
The coefficients of the chiral expansion are input parameters
to $\chi $PT, and their values are determined from experimental observables.
These coefficients therefore provide a convenient representation
of a large body of data relevant to low-energy QCD.

Quantum chromodynamics is formulated in terms of unobserved degrees of
freedom -- quarks and gluons.  The presence of these fundamental constituents
of hadrons is inferred through the analysis of deep inelastic lepton
scattering.  Nevertheless, the successful application of
effective theories such as $\chi $PT to a broad range of
low-energy strong-interaction phenomena suggests that quarks and gluons
may be replaced by local effective hadronic degrees of freedom in the
low-energy domain.  This success is largely due to the separation in the hadron
spectrum between the Goldstone modes and higher mass states.
At intermediate energies it is not clear that such a
description remains effective\cite{isgurjaffe}, nor is it clear that
explicit quark and gluon degrees of freedom are essential.  An ideal
perspective
on this problem would be provided if composite hadron fields and their
interactions could be modeled in a manageable form in terms of the point
fields of QCD.  Functional integral calculus formulates this problem as an
exercise in changing the variables of integration from quark and gluon
fields to hadron fields\cite{cahill92}.  An obvious advantage of this
approach is that the effective hadron-field interactions retain knowledge
of their subhadronic origin.

The notion that such a change of variables exists for QCD in the low energy
domain is implicit in the success of the above mentioned hadronic formulations.
The explicit operation of changing variables allows the underlying dynamics of
the microscopic description to influence interactions at the macroscopic level.
The goal of this ``matching" program is then to perform the appropriate change
of integration variables in the functional integral formulation of QCD;
\begin{eqnarray}
\int D\bar{q}DqDA &&  \mbox{exp}\left( -S[\bar{q},q,A]\right)  \nonumber \\
 && =\int D\pi ...D\bar{N}DN...\mbox{exp}\left( -S'[\pi
,...,\bar{N},N,...]\right)
.\label{1.3}
\end{eqnarray}
Significant progress toward this goal has recently been
achieved\cite{cahill92}.

As the local integration variables in (\ref{1.3}) are identified with the  bare
hadron fields, their effective interactions are simultaneously defined.  This
process is the result of an expansion about the chiral symmetry breaking
ground state\cite{cahill85}, and an allocation of internal and center-of-mass
dynamics. The latter is prescribed by the normal-mode expansion of the free
kinetic operator of the composite particle in a manner analogous to the
interaction picture of standard quantum field theory\cite{kleinert78,cahill89}.
The tree-level effective interactions thereby obtained occur through a
dynamically regulated ``constituent-quark" loop and thus reflect the
underlying description.  The low-momentum (gradient) expansion of these
tree-level nonlocalities produce finite coefficients, and
for the Nambu-Goldstone modes is structurally consistent with $\chi $PT.

The direct derivation of the chiral coefficients from
QCD is presently inaccessible.  However, they can be derived from a class of
chiral invariant quark-based field-theory models of QCD which are
distinguished
by the form of the quark-quark interaction.  In this investigation the
sensitivity of the chiral coefficients
to the underlying quark-quark interaction is tested for a
variety of forms to determine their utility in constraining these models.
Previous work\cite{roberts88,roberts94} has demonstrated how these techniques
can be used to extract the second-order coefficients and those at fourth
order associated with $\pi - \pi $ scattering.
In the following we extend the previous work
by calculating the chiral coefficients $L_1$--$L_8$,
and further by
investigating the sensitivity of these coefficients to the infrared form of
the quark-quark interaction.  The coefficients
$L_9$ and $L_{10}$ are left for a future investigation,  however
work in that direction has been initiated\cite{frank94a}.

We find that in general, for momentum-space interactions
of the form $\alpha _s(q^2)/q^2$, the reproduction
of the accepted values of the chiral coefficients
requires the running coupling, $\alpha_s(q^2)$, to have a sufficiently large
integrated strength to produce dynamical
chiral symmetry breaking(D$\chi $SB), but is otherwise not acutely dependent
on the detailed form.  This constraint is
implemented here by fixing the
value of the pion decay constant, $f_{\pi}$, which determines the
overall scale.  The low-momentum strength of
$\alpha _s(q^2)$ is then implied by the scale at which the
infrared phenomenology is matched to the known ultraviolet form.  This scale is
allowed to vary to investigate the sensitivity of the coefficients,
and several two-parameter models for $\alpha _s(q^2)$ are employed.

We also find in particular that
the coefficients $L_5$ and $L_8$ are most sensitive to the
form of the interaction in the infrared.  These coefficients
are primarily responsible for
distinguishing the pion, kaon, and eta decay constants
and providing higher order corrections
to their masses\cite{chipt,holstein}.  The sensitivity of these
mass dependent coefficients is an indication that the hadron
spectrum is playing a role in the determination of the quark-quark interaction.
This result is consistent with a
previous investigation which shows that the
convergence radius of the chiral expansion in the current quark mass alone
strongly depends on the form of the quark-quark interaction
in the infrared region\cite{hatsuda,meissner95}.
In particular it was found there that the running coupling has to be strong in
the infrared
region in order to obtain convergence of the chiral series in the strange quark
sector.

Finally, we find that a pattern in the coefficients
emerges at tree level which is consistent with the
large $N_c$ expansion in QCD and can be traced to approximations that are
made to QCD here in deriving the low-energy expansion\footnote{The role of the
$\eta_0$ and the associated anomaly are neglected here in considering the
$N_c$ dependence of the chiral coefficients.  This
question was first addressed in Ref.\cite{witten} and is reviewed in
Ref.\cite{holstein}.}.
In this
way the consequences of the model assumptions can be
directly observed.
The present investigation further
provides a significant reduction in the number of parameters
needed to represent the low-energy QCD data mentioned above, and thereby
deepens
our understanding of low-energy QCD.

The paper is organized as follows.  In section II the path from QCD
to $\chi $PT is explored.  In Section III the consequences of
the approximations made in Section II, along with the model dependence of the
results, are investigated.  Finally conclusions are offered
in Section IV.

\section{QCD, the effective quark-quark interaction, and $\chi $PT}

\subsection{From QCD to the effective quark-quark interaction}

A global color symmetry model(GCM)\cite{cahill85,roberts88} that is based
upon an effective
quark-quark interaction can be defined through a truncation of QCD as follows.
The generating functional for QCD in the Euclidean metric is given by
\begin{equation}
Z[\psi ,\bar{\eta },\eta ]=\int D\bar{q}DqDA \; \mbox{exp}
\left( -S[\bar{q},q,A^a_{\mu }]-\bar{q}\psi q+ \bar{\eta}q+\bar{q}\eta
\right) \label{2.1}
\end{equation}
 and can be rewritten as
\begin{equation}
Z[\psi ,\bar{\eta },\eta ]\!=\! \! \int \! D\bar{q}Dq \; \mbox{exp}\!
\left[ -\!\int \bar{q}(\not \! \partial +\psi )q +\bar{\eta }q+\bar{q}
\eta \right] \mbox{exp}\! \left( W\! \left[ ig\bar{q}\frac{\lambda ^a}
{2}\gamma _{\mu }q\right] \right) \label{2.2}
\end{equation}
with $W[J]$ given by
$\mbox{exp}\left( W[J]\right)=\int DA \; \mbox{exp}\left( -\frac{1}{4}
F^a_{\mu \nu }F^a_{\mu \nu }+J_{\mu }^aA_{\mu }^a\right) $.  Here $\bar{\eta
}$,
$\eta $ and $\psi $ are external source fields.

The quantity $W[J]$ has an expansion in gluon $n$-point functions
starting at second order;
\begin{equation}
W[J]=\frac{1}{2}\int D^{ab}_{\mu \nu }(x,y)J_{\mu }^a(x)J_{\nu
}^b(y)+W_R[J],\label{2.3}
\end{equation}
where $W_R[J]$ involves gluon $n(\ge 3)$-point functions.  It is worth
noting that the $n$-point functions have mass dimension $[mass]^n$.
One might
therefore hope that for low-energy hadron physics the low-dimension functions
would provide a good description.

By replacing the quark field variables in $W_R[J]$ by their
source derivatives, the generating functional of QCD can be written as
\begin{equation}
Z[\psi ,\bar{\eta },\eta ]= \mbox{exp} \left( W_R\left[ ig\frac{\delta }
{\delta \eta }\frac{\lambda ^a}{2}\gamma _{\mu }\frac{\delta }
{\delta \bar{\eta }}\right] \right) Z_{GCM}[\psi ,\bar{\eta },\eta ]\label{2.4}
\end{equation}
where $Z_{GCM}[\psi ,\bar{\eta },\eta ]\equiv \int D\bar{q}Dq \; \mbox{exp}
\left( -S_{GCM}[\psi,\bar{q},q]+\bar{\eta }q+\bar{q}\eta \right) $ with
\begin{eqnarray}
S_{GCM}[\psi ,\bar{q},q]\equiv \int d^{4}xd^{4}y\Biggl\{ \bar{q}(x)\Bigl[
\bigl( \not \! \partial _{x} +\psi (x)&&  \bigr) \delta(x-y)\Bigr]
q(y)\nonumber \\
 && + \frac{g^{2}}{2}j_{\nu
}^{a}(x)D(x-y)j_{\nu }^{a}(y)\Biggr\} .\label{2.6}
\end{eqnarray}
Here $j_{\nu }^{a}(x)\equiv \bar{q}(x)\frac{\lambda ^a}{2}\gamma _{\nu }q(x)$
is the quark color current, and for convenience a gauge for the gluon
propagator
$D^{ab}_{\mu \nu }(x-y)=\delta _{ab}\delta _{\mu \nu }D(x-y)$ is employed. From
here
forward we work within the model truncation defined in (\ref{2.6}).

The primary benefit of this truncation is that a reasonably
solvable model is obtained,
which is nevertheless sufficiently general to address a variety of
phenomenological issues such as the role of quark-quark interactions
in effective hadronic field theories.  This model as well
maintains the global symmetries of QCD and permits a $1/N_c$ expansion.

The primary loss of working at
this
level is that of the local color gauge invariance of QCD.
The consequences of
this loss are unclear, but are determined by the operation of
$W_R$ in (\ref{2.4}).  The approximation of a local symmetry by
a global symmetry
is similar to the approximation of general relativity by special relativity.
If the relevant field is sufficiently weak in the region of interest,
then such an approximation is reasonable.
In the case of localized color-singlet states one might hope that color
neutrality could provide such a scenario\cite{frank95c}.

It should be noted that the Nambu--Jona-Lasinio(NJL) model\cite{nambu61} is
obtained from
(\ref{2.6}) in the limit $D(x-y)=\delta (x-y)/{\cal M}^2$,
with ${\cal M}$ the appropriate
mass scale.  The chiral coefficients in the NJL model have
been investigated \cite{ruizarriola,bijnens,klevansky}, and are a limiting
case of the present investigation.  Our interest here is with the more general
question of the model dependence of these coefficients.  The present
description
also allows the discussion of higher mass excitations due to the nonlocal
interaction.

\subsection{From the quark-quark interaction to $\chi $PT}

\subsubsection{Bosonization and saddle-point expansion}

The meson sector of the variable change implied in (\ref{1.3}) is
revealed by first identifying field combinations (currents) with the
transformation properties of mesons.  This is achieved through a
Fierz reordering of the current-current term of the action (\ref{2.6})
to obtain
\begin{equation}
\frac{g^2}{2}j_{\mu }^a(x)D(x-y)j_{\mu }^a(y)=-\frac{g^2}{2}J^{\theta
}(x,y)D(x-y)J^{\theta
}(y,x),\label{2.7}
\end{equation}
where $J^{\theta }(x,y)\equiv \bar{q}(x)\Lambda ^{\theta }q(y)$ and
the minus sign in (\ref{2.7}) arises from the Grassmann nature
of the quark field variables.  Here the quantity $\Lambda ^{\theta }$
is the direct product of Dirac, flavor $SU(3)$  and color matrices;
\begin{equation}
\Lambda ^{\theta }=\frac{1}{2}\left( {\bf 1}_{D},i\gamma
_{5},\frac{i}{\sqrt{2}}\gamma _{\nu },\frac{i}{\sqrt{2}}\gamma _{\nu
}\gamma _{5}\right) \otimes \left( \frac{1}{\sqrt{3}}{\bf
1}_{F},\frac{1}{\sqrt{2}}\lambda _F^a\right) \otimes \left( \frac{4}{3}{\bf
1}_{C},\frac{i}{\sqrt{3}}\lambda _C^{a}\right) ,\label{2.8}
\end{equation}
which contains, in particular, color singlet $q\bar{q}$ combinations.  It
should be noted, however, that there are also color octet $q\bar{q}$
combinations present in (\ref{2.8}).  An alternate color Fierz reordering,
\begin{equation}
\sum _{a=1}^{8}\left( \lambda _a\right) _{ij}\left( \lambda _a\right)
_{kl}=\frac{4}{3}\delta
_{il}\delta _{kj}+\frac{2}{3}\sum_{m=1}^{3}\epsilon _{mik}\epsilon
_{mlj},\label{2.9}
\end{equation}
eliminates the color octet $q\bar{q}$ sector in favor of color
triplet-antitriplet $qq$ combinations and leads naturally to
baryons\cite{cahill92}.  This alternate approach, although natural
for the investigation of baryons, is unnecessary for the investigation
of meson interactions of interest here.  The interested reader is encouraged
to consult Ref.\cite{cahill92} and references therein for details of the
baryon sector.

Having identified field combinations with the transformation properties of
mesons, the current-current term of the action (\ref{2.6}) is eliminated
by multiplying the partition function by unity in the Gaussian form
\begin{equation}
1={\cal N}\int D{\cal B}\; \mbox{exp}\! \left[ -\int d^4xd^4y
\frac{{\cal B}^{\theta }(x,y){\cal B}^{\theta }(y,x)}{2g^2D(x-y)}\right]
\label{2.10}
\end{equation}
and shifting the bilocal-field integration variables as
${\cal B}^{\theta }(x,y)\rightarrow {\cal B}^{\theta }
(x,y)+g^2D(x-y)J^{\theta }(y,x)$\cite{shrauner}.  This requires in
particular
that the bilocal fields ${\cal B}^{\theta }(x,y)$ display the same
symmetry transformations  as the bilocal currents
$J^{\theta }(y,x)$\cite{frank94a}.

The partition function now has the form
$Z[\psi]={\cal N}\int D{\cal B}D\bar{q}Dq\; e^{-S[\psi,\bar{q},q,{\cal B}]} $
where
\begin{equation}
S[\psi, \bar{q},q,{\cal B}]=\! \! \int d^4xd^4y\; \bar{q}(x)\!
\left[ (\not \! \partial _x +\psi(x) )\delta (x\! -\! y)+\Lambda ^{\theta }
{\cal B}^{\theta }(x,y)\right] \! q(y) + \frac{{\cal B}^{\theta }(x,y)
{\cal B}^{\theta }(y,x)}{2g^2D(x-y)}.\label{2.12}
\end{equation}
The action (\ref{2.12}) is quadratic in the quark fields which allows
the Grassmann integration to be performed by standard methods.  The
resulting expression for the partition function in terms of the
bilocal-field integration is $Z[\psi]={\cal N}\int D{\cal B}\;
e^{-S[\psi,{\cal B}]}$ where the action is given by
\begin{equation}
S[\psi,{\cal B}]=-\mbox{TrLn}\left[ G^{-1}\right] +\int d^4xd^4y \frac{{\cal
B}^{\theta
}(x,y){\cal B}^{\theta }(y,x)}{2g^2D(x-y)},\label{2.13}
\end{equation}
and the quark inverse Green's function, $G^{-1}$, is defined as
\begin{equation}
G^{-1}(x,y)\equiv (\not \! \partial _x+\psi(x) )
\delta (x\! -\! y)+\Lambda ^{\theta }{\cal B}^{\theta }(x,y).\label{gf}
\end{equation}

This replacement of the quark-field integration with the bilocal-field
integration represents an exact functional change of variables.
Observable quantities extracted from the partition function are unaffected
by the variable change, but are now  expressed in terms of effective
(meson) degrees of freedom.  A benefit of this is that the effective
mesonic interactions, which are generated from the quark-field
determinant in (\ref{2.13}), represent a summation of quark processes,
and are easily exposed by expanding in powers of the bilocal fields.
The structure of these interactions is illustrated in Fig.1.
At this level the bilocal fields interact through a bare quark
loop as in Fig.1a, and do not readily display the dynamics expected
of quark bound states of QCD.  However, as the notion of bare mesons
is developed, this picture of their interactions is simultaneously refined.

In anticipation of dynamical chiral symmetry breaking, bare mesons are
defined in terms of the fluctuations about the saddle point of the action
(which is equivalent to the classical vacuum).  This choice of an
expansion point harbors profound dynamical consequences in that it
largely determines both the structure and interactions of the bare
mesons.  In particular, this choice leads to the
rainbow Dyson-Schwinger equation of the quark self energy, and the ladder
Bethe-Salpeter equation for the internal structure of the bare mesons.  More
importantly, as a result of grouping this particular class of diagrams into
bare mesons, the expansion about the classical vacuum leads to results for the
chiral coefficients which are consistent with large $N_c$ QCD,
as is discussed in Section III.

The saddle-point of the action is defined as
$\left. \frac{\delta S}{\delta {\cal B}}\right| _{{\cal B}_0,\psi =0}=0$ and
is given by
\begin{equation}
{\cal B}^{\theta }_0(x-y)=g^2D(x-y)\mbox{tr}\left[ \Lambda ^{\theta
}G_{0}(x-y)\right]
.\label{spoint}
\end{equation}
These configurations are related to nonlocal vacuum condensates\cite{gsm}
and provide self-energy dressing of the quarks through the definition
$\Sigma (p)\equiv \Lambda ^{\theta }{\cal B}^{\theta }_0(p)=
i\not \! p\left[ A(p^2)-1\right] +B(p^2)$, where
\begin{equation}
\left[ A(p^{2})-1\right] p^{2} =g^{2} \frac{8}{3}\int \frac{d^{4}q}{(2\pi
)^{4}}D(p-q)\frac{A(q^{2})q\cdot p}{q^{2}A^{2}(q^{2})+B^{2}(q^{2})},
\label{aeq}
\end{equation}
and
\begin{equation}
B(p^{2})=g^{2}\frac{16}{3}\int \frac{d^{4}q}{(2\pi
)^{4}}D(p-q)\frac{B(q^{2})}{q^{2}A^{2}(q^{2})+B^{2}(q^{2})}
.\label{beq}
\end{equation}
This dressing comprises the
notion of ``constituent" quarks by providing a mass
\begin{equation}
M(p^2)=\frac{B(p^2)}{A(p^2)}.\label{mass}
\end{equation}
Their role as constituents is further displayed
by expanding the bilocal fields about the saddle point,
\begin{equation}
{\cal B}^{\theta }(x,y)=
{\cal B}^{\theta }_0(x-y)+\hat{\cal B}^{\theta }(x,y), \label{fluct}
\end{equation}
then examining the effective interactions of
the fluctuations, $\hat{\cal B}$.  These interactions are produced by
the quark-field determinant $\mbox{TrLn}\left( \not \! \partial +
\Sigma +\Lambda ^{\theta }\hat{\cal B}^{\theta }\right) $, as is
illustrated
in Fig.1b.  There it is seen that the fluctuation-field interactions now
occur through the constituent-quark loops.

The connection between the bilocal fluctuation fields and the local fields
of standard hadronic field-theory phenomenology remains to be shown.  The
bilocal fields contain information about internal excitations of the
$q\bar{q}$ pair in addition to their net collective or center-of-mass motion
which is to be associated with the usual local field variables.  A separation
of the internal and center-of-mass dynamics is achieved by considering the
normal modes of the free kinetic operator of the bilocal fields in a manner
which is analogous to the interaction representation of standard quantum
field theory.  Details of the localization procedure
can be found in Refs.\cite{cahill89} and \cite{frank95b}.
The process amounts to a projection of the
bilocal field $\hat{\cal B}^{\theta }$ onto a complete set of internal
excitations $\Gamma _{n}^{\theta }$ with the
remaining center-of-mass degree of freedom represented by the coefficients
$\phi ^{\theta }_{n}(P)\equiv \int d^4q\hat{\cal B}^{\theta }(P,q)
\Gamma _{n}^{\theta }(P,q)$.  The bilocal fluctuations can thus be written as
\begin{equation}
\hat{\cal B}^{\theta }(P,q)=\sum _n\phi ^{\theta }_{n}(P)
\Gamma _{n}^{\theta }(P,q).
\label{lfields}
\end{equation}

The functions $\Gamma _{n}^{\theta }$ are in general eigenfunctions of the
the free kinetic operator of the bilocal fields.  At the mass shell point,
$P^2=-M_n^2$, they satisfy the homogeneous Bethe-Salpeter equation
in the ladder approximation for the
given quantum numbers $\theta $ and mode $n$.  This modal expansion is
then used to localize the action.

At tree level the local fields $\phi _{n}^{\theta }$ interact through a
dynamically regulated constituent-quark loop, as is illustrated, for
example, in Fig.2.  These ``effective interactions" thus reflect
the underlying QCD structure.  The  intrinsic nonlocality plays a dual
role in the subsequent description of physical phenomena.  First, when
sufficiently short length scales are probed as in the large momentum
behavior of hadronic form factors, the nonlocal structure is directly
observed\cite{frank94a,formfactors}.  Second, independent of external probes,
the nonlocality
provides a regulation of internal loop integrations, and serves to
suppress hadron-loop effects\cite{hollenberg92alkofer93}.
The present approach can also accommodate
extensions of low-energy effective theories through the consideration
of the higher mass states,  and therefore provides a consistent
framework in which many of the issues facing hadronic field theories
might be addressed.

\subsubsection{Dynamical chiral symmetry breaking}

In the following discussion, D$\chi $SB is associated with the occurance
of a massless Goldstone mode that is related to the dynamical generation
of a scalar amplitude in the quark self energy in the limit of vanishing
quark mass.  We begin by considering the axial-vector Ward identity
in the chiral limit given by\cite{delbourgo79,review}
\begin{equation}
\left. P_{\mu }\Gamma _{\mu }^5(P,q)\right| _{m=0}=G^{-1}(q+P/2)
\gamma _5 +\gamma _5G^{-1}(q-P/2).\label{awi}
\end{equation}
It is well known\cite{delbourgo79,review} that in the chiral limit the
axial-vector vertex contains a zero-momentum pole of the form
\begin{equation}
\Gamma _{\mu }^5(P,q){\stackrel{P\rightarrow 0}{\rightarrow }}\frac{P_{\mu}}
{P^2}\Gamma _5(0,q)f_{\pi },\label{avvcl}
\end{equation}
associated with the massless Goldstone mode.
It should be noted that in (\ref{avvcl}) the
quark-pseudoscalar vertex $\Gamma _5$ is also evaluated in the vicinity
of $P=0$, and is therefore a solution of the homogeneous
Bethe-Salpeter equation for the pseudoscalar bound state.

Operating on (\ref{avvcl}) with $P_{\mu }$ and comparing with (\ref{awi})
obtains
\begin{equation}
\Gamma _5(0,q)=2\gamma _5\frac{B(q^2)}{f_{\pi }}.\label{gtr}
\end{equation}
This is the Goldberger-Treiman relation for the quark-pseudoscalar vertex.
The fact that the quark self-energy function $B$
occurs as the
residue of the zero momentum pole in the quark-axial-vector vertex is
equivalent to a statement of Goldstone's theorem in this context.
It is also readily verified that the ladder Bethe-Salpeter equation for the
pseudo-scalar
Goldstone mode reduces to the self-energy equation
(\ref{beq})\cite{delbourgo79,review}.

Since our interest here is the effective action for the Goldstone
modes, we neglect all of the higher mass fluctuations present in the
bilocal fields.  This implies that the full bilocal field of
Eq.(\ref{fluct}) can be written, using the expressions
(\ref{spoint}) and (\ref{gtr})
for the saddle point and the Bethe-Salpeter amplitude of the Goldstone modes
respectively, as
\begin{equation}
\Lambda ^{\theta }{\cal B}^{\theta }(x,y)=\Sigma (x-y) +B(x-y)
\left[ U_5\left(\frac{x+y}{2}\right) -1\right] ,\label{btheta}
\end{equation}
where $U_5(x)=P_RU(x)+P_LU^{\dagger }(x)$ with $P_{R,L}$ the standard
right-left projection operators.  For the $SU(3)$ flavor case under
consideration here the chiral field $U$ is defined as
$U\equiv e^{i\lambda ^a\phi ^a/f_{\pi }}$.
It should be stressed at this point that we have not integrated over the
higher mass states, but have simply neglected them.  The effect of
including and integrating over the higher mass states is addressed in
Section III.

\subsubsection{The low-energy expansion}

For the application to low-energy observables, an expansion of the
action to fourth order is now considered.  The usual chiral power
counting is observed\cite{chipt,holstein}.
In order to preserve the chiral invariance of the full action (\ref{2.6}),
the quarks have to be coupled to the external source field $\psi (x)$,
which transforms in a certain way under chiral rotations\cite{chipt,holstein}.
In performing the gradient expansion it is important to keep the $x$ dependence
of this external field.
After carrying out the gradient expansion to fourth order, we will employ
the equation of motion which is obtained at second order
and depends on the external field $\psi (x)$, and
then finally we will identify $\psi(x)$ with the current quark mass matrix.
Failure to keep the $x$ dependence of $\psi$ to the very end violates
chiral invariance and will render unphysical results for some
of the low-energy coefficients.
This approach differs somewhat from the previous work of
Refs.\cite{roberts88} and \cite{roberts94} where the equation
of motion is not employed.  However, there the mass-dependent
fourth-order coefficients are not considered.

We consider here only the real
contribution to the effective action.  The imaginary contribution, which
contains the Wess-Zumino term, has also been investigated in
Refs.\cite{roberts88} and \cite{roberts94}, and the interested reader is
encouraged to consult these references for more details.

The restriction of the fluctuations to Goldstone modes with
$UU^{\dagger }=1$, as in
(\ref{btheta}), entails that the second term of the action in
Eq.(\ref{2.13}) is independent of the fields $U$ and can therefore
be neglected.  The real contribution to the action is then given by
\begin{equation}
{\cal S}\equiv \mbox{Re}[S]=-\frac{1}{2}\mbox{TrLn}
\left( G^{-1}\left[ G^{-1}\right] ^{\dagger }\right) ,\label{res}
\end{equation}
where $G^{-1}$ is, from (\ref{gf}) and (\ref{btheta}), given by
\begin{equation}
G^{-1}(x,y)=\gamma \cdot \partial _xA(x-y)+\psi \left(
\frac{x+y}{2}\right) \delta(x-y) +B(x-y)U_5\left(
\frac{x+y}{2}\right) .\label{gfu}
\end{equation}
By expanding the logarithm and dropping irrelevant constant terms,
Eq.(\ref{res}) can further be written as
\begin{equation}
{\cal S}=\frac{1}{2}\sum _{n=1}^{\infty}\frac{(-1)^n}{n}
\mbox{Tr}\left( a+b+c+d\right) ^n ,\label{ssum}
\end{equation}
where $a$, $b$, $c$, and $d$ are non-commuting operators formed
from $A$, $B$, $U_5$, and $\psi $, and are at least of order one,
one, two, and three in chiral counting respectively.
The explicit form of these
operators is given in the appendix.

The effective chiral action to the desired order is now obtained
by expanding the sum in Eq.(\ref{ssum}) and expanding the operators
$a$--$d$ in gradients.  The result to fourth order is (in Euclidean space)
\begin{eqnarray}
{\cal S}=\int d^4x &\Biggl\{ & \frac{f^2_{\pi }}{4}\mbox{tr}
\left[ (\partial _{\mu}U)(\partial _{\mu}U^{\dagger})\right]
-\frac{f^2_{\pi }}{4}\left[ U\chi ^{\dagger} +\chi U^{\dagger}\right]
\nonumber \\
&-&L_1\left( \mbox{tr}\left[ (\partial _{\mu}U)
(\partial _{\mu}U^{\dagger})\right] \right) ^2 -L_2
\mbox{tr}\left[ (\partial _{\mu}U)(\partial _{\nu}U^{\dagger})\right]
\cdot \mbox{tr}\left[ (\partial _{\mu}U)(\partial _{\nu}
U^{\dagger})\right] \label{chiact} \\
&-&L_3\mbox{tr}\left[ (\partial _{\mu}U)(\partial _{\mu}U^{\dagger})
(\partial _{\nu}U)(\partial _{\nu}U^{\dagger})\right] +L_5
\mbox{tr}\left[ (\partial _{\mu}U)(\partial _{\mu}U^{\dagger})
(U\chi ^{\dagger} +\chi U^{\dagger})\right] \nonumber \\
&-&L_8\mbox{tr}\left[ \chi U^{\dagger}\chi U^{\dagger} +
U\chi ^{\dagger }U\chi ^{\dagger }\right] \Biggr\} ,\nonumber
\end{eqnarray}
where $\chi (x)=-2\langle \bar{q}q\rangle\psi (x)/f^2_{\pi }$ and the
remaining trace is over flavor.
In obtaining this result the equation of motion
\begin{equation}
(\partial ^2U)U^{\dagger}+(\partial _{\mu}U)(\partial _{\mu}U^{\dagger})
+\frac{1}{2}(\chi U^{\dagger }-U\chi ^{\dagger })=0\label{eqmo}
\end{equation}
and the $SU(3)$ relation\cite{holstein}
\begin{eqnarray}
\mbox{tr}\left[ (\partial _{\mu}U)(\partial _{\nu}U^{\dagger})
(\partial _{\mu}U)(\partial _{\nu}U^{\dagger})\right]& = &
\frac{1}{2}\left( \mbox{tr}\left[ (\partial _{\mu}U)
(\partial _{\mu}U^{\dagger})\right] \right) ^2 \nonumber \\
+\mbox{tr}\left[ (\partial _{\mu}U)(\partial _{\nu}U^{\dagger})\right]
&\cdot &\mbox{tr}\left[ (\partial _{\mu}U)(\partial _{\nu}
U^{\dagger})\right] - 2\mbox{tr}\left[ (\partial _{\mu}U)
(\partial _{\mu}U^{\dagger})
(\partial _{\nu}U)(\partial _{\nu}U^{\dagger})\right] \label{su3}
\end{eqnarray}
have been used.  Explicit forms of the coefficients are given in the appendix.

\section{Results and Discussion}

Several conclusions can be drawn directly from the low-energy expansion
(\ref{chiact}).  It is immediately apparent that the coefficients $L_4$,
$L_6$, and $L_7$ vanish.  It is also evident, by application of the $SU(3)$
relation (\ref{su3}) (see appendix), that $L_2=2L_1$.  These relationships
are expected in the large $N_c$ limit\footnote{In the presence of
the $U_A(1)$ anomaly the coefficient $L_7$ is of order $N_c^2$
\cite{witten,holstein}.  Our neglect of the $\eta _0$ here leads
to the vanishing of $L_7$.  }
of QCD\cite{holstein}.
The fact that they are produced here is perhaps not too surprising
and can be linked to our truncation of the QCD action to include only
the gluon two-point function.

With only a two-point quark-quark interaction,
the large $N_c$ limit leads to a description of mesons as a sum of
ladder exchanges.  Our description of mesons as fluctuations about
the saddle point of the action is equivalent to the $1/N_c$ expansion and,
in this model truncation, leads directly to the ladder approximation.
Our further neglect of
higher mass states, explicitly excludes intermediate states of pure
glue which are ``$N_c$ suppressed".
A departure from this tree-level pattern in the
coefficients would therefore have to arise
in the present formalism by including and integrating over the higher mass
mesons, which we have explicitly excluded in Eq.(\ref{btheta}).
The role of the underlying description is thus
clearly displayed in the pattern of the chiral coefficients.

Examples of the diagrams that are generated by integrating over
higher mass mesons are illustrated in Fig.3.  The diagram
of Fig.3a is of order one in $N_c$ counting and produces
departures from the large $N_c$ relations, while the diagram of Fig.3b
is of order $N_c$ and produces, for example, the $\rho $-pole in
$\pi $--$\pi $ scattering.  All of the contributions that we
are presently considering are of order $N_c$ and arise from a
single quark loop\footnote{The double trace terms proportional
to $L_1$ and $L_2$ in (\ref{chiact}) arise here from the $SU(3)$
relation (\ref{su3}) and originate from a single quark loop.}.

The remaining nonzero coefficients must be evaluated numerically.  These
depend explicitly of the values of the self-energy functions $A$ and $B$
in Eqs.(\ref{aeq}) and (\ref{beq}) respectively, and
are therefore implicitly dependent on the quark-quark interaction $D$.
The procedure is then to select a form for the function $D$, solve the
coupled nonlinear equations (\ref{aeq}) and (\ref{beq}) for $A$ and $B$
respectively,
and then evaluate the pion decay constant $f_\pi$, the condensate
$\langle \bar{q}q \rangle $, and the coefficients $L_1$, $L_3$, $L_5$,
and $L_8$.

The quark-quark interaction $D$ has the form
\begin{equation}
g^2D(s)=\frac{4\pi \alpha (s)}{s} ,\label{dofq}
\end{equation}
where $s=q^2$, and we investigate three different two-parameter models
for $\alpha (s)$;
\begin{eqnarray}
\alpha _1(s)&=&3\pi s\chi ^2\frac{e^{-s/\Delta}}{4\Delta^2}
+\frac{\pi d}{\mbox{ln}(s/\Lambda^2 +e)}\nonumber \\
\alpha _2(s)&=&\pi d\left[ \frac{s\chi^2}{s^2+\Delta }+
\frac{1}{\mbox{ln}(s/\Lambda^2 +e)}\right] \label{alfs} \\
\alpha _3(s)&=&\pi d\left[ \frac{1+\chi e^{-s/\Delta}}
{\mbox{ln}(s/\Lambda^2 +e)}\right] .\nonumber
\end{eqnarray}
Each of these forms incorporates the one-loop perturbative result
for large $s$ (here $\Lambda =0.2$GeV and $d=12/(33-2N_F)=12/27$),
and extrapolates differently into the
low-momentum region.  The two low-momentum parameters,
$\chi $ and $\Delta $, are varied with the pion decay constant held fixed
at $f_\pi =86$MeV.  This value is appropriate at zero-momentum
rather than the pion-mass-shell value of $93$MeV, however the
results are not very sensitive to this small difference.
By fixing $f_\pi$ the overall scale of D$\chi $SB is fixed.
The remaining independent parameter
is associated with the matching scale to the perturbative form.

The running coupling for the three cases listed in Eq.(\ref{alfs}) are plotted
in Figs. \ref{fig4}, \ref{fig5}, and \ref{fig6} along with the
corresponding solutions of Eqs.(\ref{aeq}) and (\ref{beq}) for the
self-energy functions.  In all three cases as the matching point to the
perturbative form is decreased to lower momentum, the infrared strength
must be increased to maintain the fixed value of $f_\pi$.  Thus the
integrated strength of $\alpha $ is largely constant.  This trend is also
present to a lesser extent in the self-energy functions.

The first model, $\alpha _1$ in Eq.(\ref{alfs}), has been used in previous
investigations of the present type\cite{praschifka89}.  There the parameters
were fixed at $\Delta =0.002 $GeV$^2$ and $\chi =1.14$GeV, which leads to
a slightly lower value of $f_\pi $.  The small $\Delta $ limit of this
model obtains a matching point near zero momentum and a delta-function behavior
in the quark-quark interaction $D$.  This limit has been used previously to
model confinement\cite{review}.
The infrared contribution to the second model, $\alpha _2$, generates a
$1/q^4$ singularity in the quark-quark interaction $D$ in the limit as
$\Delta \rightarrow 0$.  Such a singularity has also
been considered previously as a model of confinement\cite{marciano78}.
The $1/q^4$ form falls much slower than the Gaussian in $\alpha _1$, and hence
leads to much higher matching scales.
Finally the third model $\alpha _3$ has been chosen here to be
structurally different from $\alpha _1$ and $\alpha _2$ in order to further
illustrate the independence of the results to the details of the low-momentum
parameterization.
The corresponding results for the low energy coefficients $L_1$, $L_3$,
$L_5$, and $L_8$
are displayed in Tables I-III, respectively.

In all of the three cases the same pattern is observed:
The coefficients $L_1$ and $L_3$, which are responsible for $\pi$-$\pi$ and
$K$-$K$
scattering, are nearly independent of the form of $\alpha (s)$ and therefore
on the form of the quark-quark interaction, provided that the integrated
strength
of $\alpha (s)$ is fixed by $f_\pi$.
On the other hand, the mass dependent coefficients, $L_5$ and $L_8$,
are more strongly
dependent on the actual form of the interaction.
For example with $L_5$,  in order to
reproduce the experimental value, forms of $\alpha (s)$
with a small matching scale, i.e. which are relatively strong in the infrared
region, are
required.
This observation is in coincidence with the result of Ref.\cite{meissner95},
where it is shown that quark-quark interactions with a low matching
scale are also required
to achieve convergence of the chiral series in the strange quark sector.
Furthermore an explanation for the success of the
``delta-function-plus-tail" type models (obtained for example from $\alpha_1$
in the limit $\Delta \to 0$) in describing chiral
observables\cite{frank96a} is offered by this fact.

We also find that the results for the fourth-order coefficients are rather
insensitive to the asymptotic UV tale of $\alpha (s)$;
even omitting this tail completely gives no significant changes,
again provided that $f_\pi$ is fixed.

An increased accuracy in the experimental
determination  of the coefficients would make tighter restrictions
on the quark-quark interaction, however the additional investigation of
higher mass excitations is clearly required to gain
detailed information on its infrared form\cite{frank95a}.

It has frequently been stated with regard to the fourth order coefficients
that QCD ``seems to predict that deviations from the lowest order
chiral relation must be in such a form as to reproduce
the low energy tails of the light resonances, in particular the
$\rho $."\cite{holstein}.
Here we have explicitly neglected in Eq.(\ref{btheta}) the $q\bar{q}$
fluctuation associated with the $\rho $, and illustrated in Fig.3 how
the $\rho $-pole contribution would arise.  One might then ask: What is the
mechanism that produces the $\rho $-tail-like contribution to the
coefficients here?

This question is easily answered by again considering the diagram of Fig.2.
There is a $q\bar{q}$ pair in the intermediate state which arises
from the quark loop structure of the interactions.  The integrands of
these quark loops are peaked at a momentum $q_{peak}$ such that
the constituent mass of Eq.(\ref{mass}) gives
\begin{equation}
M(q^2)\rightarrow M(q^2_{peak})\approx 300-400\mbox{MeV}.
\end{equation}
This $q\bar{q}$ pair can have the quantum numbers of the $\rho $,
and carries sufficient mass to contribute the $\rho $-tail effect
away from the pole.

\section{Conclusions}

We have made a detailed examination of the low-energy chiral expansion
from the standpoint of the model truncation of QCD given in Eq.(\ref{2.6}).
The structure of the model maintains the global symmetries of QCD
(including global color symmetry), and permits a $1/N_c$ expansion.
The infrared momentum dependence of the quark-quark interaction
is phenomenological input to the model; here three different two-parameter
forms are investigated.

We find that by truncating QCD to include only a two-point
quark-quark interaction and describing mesons as fluctuations
about the saddle point of the effective action, one obtains a pattern
in the chiral coefficients which is consistent with large $N_c$ results in
QCD.  This conclusion can be understood by considering the
$1/N_c$ expansion within the model truncation, and provides a direct link
between the model assumptions and consequences for physical observables in
QCD, independent of the phenomenological treatment of the
quark-quark interaction.  The structure of the underlying theory is
in this way displayed by the pattern in the chiral coefficients.  The
departure from the large $N_c$ result is provided here by the
integration over higher mass states.

We find that by fixing the pion decay constant,
an integrated strength
of the running coupling is prescribed.  This sets the scale for D$\chi $SB.
The remaining independent parameter is associated with the matching
scale to the perturbative form of the running coupling.  The chiral
coefficients $L_1$ and $L_3$, which are related to
$\pi$-$\pi$ and $K$-$K$ scattering data are nearly
insensitive to this scale.
It appears, therefore, that any chiral quark-quark interaction which is
capable of D$\chi $SB
can be expected to reproduce these coefficients
and the corresponding low energy meson scattering data.
However, some constraint on the matching
scale is provided by the
sensitivity of the mass-dependent coefficients
$L_5$ and $L_8$, which favor interaction forms that
are strong in the infrared domain.

Finally we conclude that the model truncation that is employed here
reproduces low-energy QCD, as represented by $\chi $PT, quite well.
More importantly, this model is not limited to low energy and might
therefore be used to extend low-energy effective theories.

\acknowledgements

This work was supported by the Department of Energy under Grant \#
DE-FG06-90ER40561,
and the National Science Foundation under Grant \# PHYS-9310124.
The authors wish to thank Barry Holstein for
suggesting the investigation of the chiral coefficients.  MRF is
grateful for helpful discussions with David Kaplan.
TM would like to thank the Institute for Nuclear Theory for
hospitality during numerous visits while this work has been completed.

\appendix
\section{}

The operators in Eq(\ref{ssum}) are defined as
\begin{eqnarray}
a&\equiv &\frac{1}{x}[\gamma _{\mu}\bar{A}_{\mu}U_5^{\dagger}B
+BU_5\gamma _{\mu }
\bar{A}^{\dagger }_{\mu}]\nonumber \\
b&\equiv &\frac{1}{x}B[U_5U_5^{\dagger }-1]B\nonumber \\
c&\equiv &\frac{1}{x}[BU_5\psi ^{\dagger}+
\psi U_5^{\dagger}B+\psi ^{\dagger }\psi ]\label{abcdx} \\
d&\equiv &\frac{1}{x}[\gamma _{\mu}\bar{A}_{\mu}\psi ^{\dagger} +
\psi \gamma _{\mu }\bar{A}^{\dagger }_{\mu}]\nonumber \\
x&\equiv &\gamma _{\mu}\bar{A}_{\mu}\gamma _{\nu }\bar{A}^{\dagger }_{\nu}
+B^2 ,\nonumber
\end{eqnarray}
where for example
\begin{eqnarray}
<x_1|\bar{A}_{\mu }|x_2>=\partial _{\mu {x_1}}A(x_1-x_2) & , &
<p_1|\bar{A}_{\mu
}|p_2>=i{p_1}_{\mu }A(p_1^{2})\delta (p_1-p_2) \nonumber \\
<x_1|B|x_2>=B(x_1-x_2) & , & <p_1|B|p_2>=B(p_1^{2})\delta (p_1-p_2) \label{2e3}
\\
<x_1|BU_5|x_2>=B(x_1-x_2)U_5\left(\frac{x_1+x_2}{2}\right) & , &
<p_1|BU_5|p_2>=\frac{1}{(2\pi
)^2}B\left(\frac{p_1+p_2}{2}\right)U_5(p_1-p_2)\nonumber \\
<x_1|U_5^{\dagger }B|x_2>=B(x_1-x_2)U_5^{\dagger
}\left(\frac{x_1+x_2}{2}\right) & , &
<p_1|U_5^{\dagger }B|p_2>
=\frac{1}{(2\pi )^2}B\left(\frac{p_1+p_2}{2}\right)U_5^{\dagger
}(p_2-p_1).\nonumber
\end{eqnarray}

{}From the discussion presented in the text, one can then obtain the
coefficients
\begin{eqnarray}
      f_\pi^2=F\int dss\Biggl( \left(\frac{B}{x}\right)^2[A^2 &+&
          sAA'+s^2(A')^2+s(B')^2]\nonumber \\
        && -\frac{BB'+\frac{s}{2}[(B')^2+BB'']}{x}\Biggr) \label{fpi2}
\end{eqnarray}
\begin{equation}
\langle \bar{q}q\rangle _{1 GeV}=-F\int ^{1 GeV^2}
ds s \frac{B}{x} \label{qbarq}
\end{equation}

\begin{eqnarray}
      L_1 &=& -\frac{1}{2}\lambda_{3}\nonumber \\
      L_2 &=& -\lambda_{3}\nonumber \\
      L_3 &=& -(\lambda_{2} - 2\lambda_{3} + \lambda_{1})\label{coef} \\
      L_5 &=& \lambda_{4} - \lambda_{6}\nonumber \\
      L_8 &=& -(\lambda_{5} - \frac{1}{4}\lambda_{1} +
                 \frac{1}{2}\lambda_{6})\nonumber
\end{eqnarray}
where

\begin{equation}
      \lambda_{1} = \int ds(\lambda_{11}+\lambda_{12}+\lambda_{13}
             +\lambda_{14}+\lambda_{15})
\end{equation}
with
\begin{eqnarray}
      \lambda_{11} &=& \frac{F}{32}  \frac{s^2  BB' }{ x^2 }
            \left( B'^2 + BB'' + \frac{1}{3}sBB''' +
          sB'B'' \right)\nonumber \\
      \lambda_{12} &=& \frac{F}{64}  s^2
          \Biggl(
          -8s(BB')^2  \frac{x''x - x'^2}{x^4}
          -8(BB')^2  \frac{x'}{ x^3}
          +2\frac{sBB'}{x^2}[BB'''-B'B'']\nonumber \\
         && +2\frac{BB'}{x^2}[3BB''-(B')^2]
           \Biggr) \nonumber \\
      \lambda_{13} &=& - \frac{Fs}{96x}
          \left(
           3[(B')^2 + BB'']
          +3s[BB''' + 3B'B'']
          +\frac{s^2 }{2} [BB'''' + 4B'B''' +3(B'')^2 ]
          \right)\nonumber \\
      \lambda_{14} &=& \frac{F}{8}  s  \left( \frac{B}{x}\right) ^2
          \left(
          \frac{3}{2}s^2A'A''+sAA''
          +\frac{1}{3}s^3  A'A'''
          +\frac{1}{6}s^2  AA'''
          +\frac{1}{2}s(A')^2 +AA'
          \right)\nonumber \\
      \lambda_{15} &=& \frac{F}{8}  s  B^2
          \left(
          \left[s^3  (A')^2 + s^2 AA'
          +\frac{1}{2}sA^2\right]
         \frac{(x')^2 - xx''}{x^4}
          -[s^2  (A')^2 + sAA' + A^2]
         \frac{x'}{x^3}
          \right)\nonumber
\end{eqnarray}


\begin{equation}
      \lambda_{2} = \int ds
         (\lambda_{21}+\lambda_{22}+\lambda_{23}+
\lambda_{24}+\lambda_{25}+\lambda_{26}+\lambda_{27}+
\lambda_{28}+\lambda_{29}+\lambda_{210})
\end{equation}
with
\begin{eqnarray}
      \lambda_{21} &=& \frac{F}{16}\frac{s}{x^2}
            \Biggl(
            (BB')^2
          + s B B'[(B')^2 + BB'']
          + \frac{s^2}{3}  [(B')^2 + BB'']^2\nonumber \\
           && + s B B'
            \left[ (B')^2 + BB'' +\frac{s}{3}BB'''
             + sB'B'' \right]
          \Biggr)\nonumber \\
      \lambda_{22} &=& -\frac{F}{32}s^2
          \left(
          -\frac{8}{3}s (BB')^2
           \frac{x''x-(x')^2} {x^4}
          + 2[BB'' - (B')^2] \frac{B  B'}{x^2}
          + \frac{2}{3}s(B  B')
             \frac{BB'''-B'B''}{x^2}
          \right)\nonumber \\
      \lambda_{23} &=& -\frac{2F}{3}s^2
          \left(
            \frac{1}{8}\left(\frac{BB'}{x}\right)^3
          + \frac{s}{4}(BB')^2 \frac{(B')^2 + BB''}{x^3}
          \right)\nonumber \\
      \lambda_{24} &=& \frac{Fs^3}{6} \left(\frac{BB'}{x}\right)^4\nonumber \\
      \lambda_{25} &=& 2Fs\left(\frac{B^2  B'}{x^2}\right)^2
          \left(
            \frac{s}{4}A^2
          + \frac{s^2}{3}  A  A'
          + \frac{s^3}{3}  (A')^2
          \right)\nonumber \\
      \lambda_{26} &=& - Fs\left(\frac{B^2  B'}{x^2}\right)^2
          \left(
             \frac{s^2}{3}  A  A'
          + \frac{s^3}{3}  (A')^2
          \right)\nonumber \\
      \lambda_{27} &=& -\frac{sF}{4}\frac{B^2 }{x^3 }
          \Biggl(
            A^2  BB'
          + \frac{s}{2}A^2  [(B')^2 + BB'']
          + sBB'[AA' + s(A')^2]\nonumber \\
          && + \frac{2}{3}s^2  [AA' + s(A')^2]
             [(B')^2 + BB'']
          \Biggr)\nonumber \\
      \lambda_{28} &=& \frac{Fs^3}{4} \frac{B^3 B'}{x^3} (A')^2\nonumber \\
      \lambda_{29} &=& \frac{sF}{4} \frac{B}{x^2}
          \Biggl(
            -\frac{4}{3}s^2  \frac{B^2 B'x'}{x^2}
            [s(A')^2 + A'A]
          - \frac{s}{2}\frac{x'B^2 B'A^2}{ x^2}
          +\frac{4}{3}s^2 B [(B')^2 + BB'']
            \frac{AA' + s(A')^2}{ x}\nonumber \\
         &&  +\frac{s}{2}\frac{BA^2 }{x}[(B')^2 + BB'']
          +\frac{B^2B'}{x}
           [s^2 ( A')^2 + sAA' + A^2]
          \Biggr)\nonumber \\
      \lambda_{210} &=& \frac{sF}{4}\left(\frac{B}{x}\right)^4
          \left(
            A^4
          + 2sA^3  A'
          + \frac{8}{3}s^2  (AA')^2
          + \frac{4}{3}s^3 A  (A')^3
          + \frac{2}{3} s^4 (A')^4
          \right)\nonumber
\end{eqnarray}


\begin{equation}
      \lambda_{3} = \int ds
          (\lambda_{31}+\lambda_{32}+\lambda_{33}+\lambda_{34}
+\lambda_{35}+\lambda_{36}+\lambda_{37}+\lambda_{38}
+\lambda_{39}+\lambda_{310})
\end{equation}
with
\begin{eqnarray}
      \lambda_{31} &=& \frac{sF}{16x^2}
            \left(
            \frac{s^2}{6}  [(B')^2 + BB'']^2
          + \frac{s}{2} B B'
            \left[ (B')^2 + BB'' +\frac{s}{3}BB'''
             + sB'B'' \right]
          \right)\nonumber \\
      \lambda_{32} &=& -\frac{s^2F}{64}
          \Biggl(
          -\frac{8}{3}s (BB')^2
           \frac{x''x-(x')^2}{x^4}
          -8(BB')^2 \frac{x'}{x^3}
          + 2[BB'' + (B')^2] \frac{B  B'}{x^2}\nonumber \\
         && + \frac{2}{3}sB  B'
             \frac{BB'''-B'B''}{x^2}
          \Biggr)\nonumber \\
      \lambda_{33} &=& -\frac{2s^2F}{3}
          \left(
            \frac{1}{4}\left(\frac{BB'}{x}\right)^3
          + \frac{s}{8}(BB')^2 \frac{(B')^2 + BB''}{x^3}
          \right)\nonumber \\
      \lambda_{34} &=& \frac{s^3F}{12} \left(\frac{BB'}{x}\right)^4\nonumber \\
      \lambda_{36} &=& - Fs\left( \frac{B^2  B'}{x^2}\right)^2
          \left(
            \frac{s}{4}A^2
          + \frac{s^2}{6}  A  A'
          + \frac{s^3}{6}  (A')^2
          \right)\nonumber \\
      \lambda_{35} &=& Fs\left(\frac{B^2  B'}{x^2}\right)^2
          \left(
            \frac{s^2}{3}  A  A'
          + \frac{s^3}{3}  (A')^2
          \right) \nonumber \\
      \lambda_{37} &=& -\frac{sF}{4}\frac{B^2}{x^3 }
          \left(
           \frac{s^2}{3}[AA' + s(A')^2]
             [(B')^2 + BB'']
          \right)\nonumber \\
      \lambda_{38} &=& -\frac{s^3 F}{4} \frac{B^3 B'}{x^3} (A')^2\nonumber \\
      \lambda_{39} &=& \frac{sF}{4} \frac{ B}{ x^2}
          \Biggl(
            -\frac{2}{3}s^2  \frac{B^2 B'x'}{x^2}
            [s(A')^2 + A'A]
          - \frac{s}{2}B^2 B'A^2 \frac{x'}{ x^2}
          +\frac{2}{3}s^2 B [(B')^2 + BB'']
            \frac{AA' + s(A')^2}{ x}\nonumber \\
        &&  +\frac{s}{2}A^2 [(B')^2 + BB''] \frac{B}{x}
          +\frac{B^2B'}{x}
           [s^2  (A')^2 + sAA' + A^2]
          \Biggr)\nonumber \\
      \lambda_{310} &=& \frac{sF}{8} \left(\frac{B}{x}\right)^4
          \left(
            -A^4
          - 2sA^3  A'
          - \frac{4}{3}s^2  (AA')^2
          + \frac{4}{3}s^3 A ( A')^3
          + \frac{2}{3}s^4 (A')^4
          \right)\nonumber
\end{eqnarray}


\begin{equation}
      \lambda_{4} =\frac{F}{2B_0} \int dss \left(
            \frac{1}{4}   \frac{ B}{x^2}
          \left(  BB'+\frac{s}{2} [(B')^2 + BB'']  \right)
          - \frac{1}{2}  \left(\frac{B}{x}\right)^3
           [  s(B')^2 + A^2 + sAA' +
              s^2  (A')^2 ]\right)
\end{equation}


\begin{equation}
      \lambda_{5} = \frac{F}{16B_0^2}\int dss\left(\frac{B}{x}\right)^2
\end{equation}

\begin{equation}
      \lambda_{6} = \frac{1}{2B_0}\int
ds(\lambda_{61}+\lambda_{62}+\lambda_{63})
\end{equation}
with
\begin{eqnarray}
      \lambda_{61} &=& -\frac{sF}{4} \frac{B}{x^2}
            [A^2 + sAA' + s^2(A')^2]\nonumber \\
      \lambda_{62} &=& \frac{sF}{8}\frac{B'+\frac{s}{2}B''}{x} \nonumber \\
      \lambda_{63} &=& - \frac{F}{4}B \left(\frac{sB'}{x}\right)^2 \nonumber
\end{eqnarray}
and finally
\begin{equation}
F\equiv\frac{4N_c}{16\pi^2}\; \; \; \; \mbox{and} \; \; \; \; B_0\equiv
-\frac{\langle \bar{q}q\rangle}{f_\pi^2}.\label{fb0}
\end{equation}

In the above  expressions,  the arguments of the functions are $s=q^2$
and the prime indicates differentiation with respect to $s$.

\begin{figure}
\caption{The effective interactions obtained from the quark determinant
are illustrated both before and after the saddle-point expansion. }
\label{fig1}
\end{figure}

\begin{figure}
\caption{An example of the effective interactions between the localized
mesons is shown.  The quark lines and vertices are dressed in the
rainbow and ladder approximations respectively. }
\label{fig2}
\end{figure}

\begin{figure}
\caption{Examples of the diagrams generated by
integrating over higher mass mesons are shown.  The diagram
in (a) is of order one in $N_c$ counting while that of (b) is of
order $N_c$.}
\label{fig3}
\end{figure}

\begin{figure}
\caption{The running coupling $\alpha _1$, and quark self-energy
functions $A$ and $B$ as shown versus $s=q^2$.  The
parameter choices maintain $f_\pi=86$MeV.}
\label{fig4}
\end{figure}

\begin{figure}
\caption{Same as Fig.4 using $\alpha _2$.}
\label{fig5}
\end{figure}

\begin{figure}
\caption{Same as Fig. 4 using $\alpha _3$.}
\label{fig6}
\end{figure}

\begin{table}[h]
\begin{center}
\begin{tabular}{|l|l|l|l|l|l|l|}
\multicolumn{7}{|c|}{$\alpha _1(s)=3\pi s\chi ^2e^{-s/\Delta}/(4\Delta^2)
+\pi d/\mbox{ln}(s/\Lambda^2 +e)$} \\ \hline
$\Delta $ (GeV$^2$) & $\chi $ (GeV) & $-\langle \bar{q}q\rangle ^{1/3}$ (MeV) &
$L_1 $(0.7$\pm $0.5) &  $L_3$(-3.6$\pm $1.3) & $L_5$(1.4$\pm $0.5) &
$L_8 $(0.9$\pm $0.3) \\ \hline
0.002 & 1.4 & 150 & 0.84 & -4.4 & 1.0 & 0.88 \\ \hline
0.02 & 1.5  & 160  & 0.82  &  -4.4  & 1.14  & 0.84  \\ \hline
0.2  & 1.65 & 173  & 0.81  &  -4.0  & 1.66  & 0.83  \\ \hline
0.4   & 1.84  & 177  & 0.80  &  -3.8  & 2.0  & 0.87 \\
\end{tabular}
\end{center}
\caption{The chiral coefficients, calculated using $\alpha _1$,
are displayed. The parameter choices listed maintain $f_\pi=86$MeV.
The ``experimental" values
in parenthesis are taken from Ref.[3].
The quark condensate
$\langle \bar{q}q\rangle$ is evaluated at 1 GeV.  }
\label{table1}
\end{table}

\begin{table}[h]
\begin{center}
\begin{tabular}{|l|l|l|l|l|l|l|}
\multicolumn{7}{|c|}{$\alpha _2(s)=\pi d(s\chi^2/(s^2+\Delta )+
1/\mbox{ln}(s/\Lambda^2 +e))$} \\ \hline
$\Delta $ (GeV$^4$) & $\chi $ (GeV)& $-\langle \bar{q}q\rangle ^{1/3}$ (MeV) &
$L_1 $(0.7$\pm $0.5) &  $L_3$(-3.6$\pm $1.3) & $L_5$(1.4$\pm $0.5) &
$L_8 $(0.9$\pm $0.3) \\ \hline
10$^{-7}$ & 0.83  & 162  & 0.82  & -4.4  & 1.28  & 0.87  \\ \hline
10$^{-4}$ & 1.02  & 167  & 0.81  & -4.2  & 1.60  & 0.91  \\ \hline
10$^{-1}$ & 1.83  & 173  & 0.79  & -3.8  & 2.36  & 1.00  \\ \hline
1         & 2.73  & 173  & 0.79  & -3.5  & 3.0  &  1.17 \\
\end{tabular}
\end{center}
\caption{Same as Table I using $\alpha _2$. }
\label{table2}
\end{table}

\begin{table}[h]
\begin{center}
\begin{tabular}{|l|l|l|l|l|l|l|}
\multicolumn{7}{|c|}{$\alpha _3(s)=\pi d(1+\chi e^{-s/\Delta})
/\mbox{ln}(s/\Lambda^2 +e)$} \\ \hline
$\Delta $ (GeV$^2$) & $\chi $ & $-\langle \bar{q}q\rangle ^{1/3}$ (MeV) &
$L_1 $(0.7$\pm $0.5) &  $L_3$(-3.6$\pm $1.3) & $L_5$(1.4$\pm $0.5) &
$L_8 $(0.9$\pm $0.3) \\ \hline
0.1   & 61.0  & 163  & 0.82  & -4.3  & 1.22  & 0.83 \\ \hline
0.4   & 24.0  & 169  & 0.81  & -4.2  & 1.48  & 0.84  \\ \hline
1.0   & 15.3  & 171  & 0.80  & -4.1  & 1.73  & 0.88  \\ \hline
2.0   & 12.2  & 172  & 0.80  & -4.0  & 1.97  & 0.95  \\
\end{tabular}
\end{center}
\caption{Same as Table I using $\alpha _3$. }
\label{table3}
\end{table}


\begin{references}

\bibitem{chipt}
J. Gasser and H. Leutwyler, Ann. Phys. (N.Y.) {\bf 158}, 142
(1983); Nucl.Phys. B {\bf 250}, 465 (1985).

\bibitem{ecker}
G. Ecker, Prog. Part. Nucl. Phys. {\bf 35}, 1 (1995), and references therein.


\bibitem{ugmeissner}
U.G. Meissner, Rep. Prog. Phys. {\bf 56}, 903 (1993);
V. Bernard, N. Kaiser and U.G. Meissner, Int. J. Mod. Phys. E {\bf 4}, 193
(1995),
and references therein.

\bibitem{isgurjaffe}
N. Isgur, in {\it From Fundamental Fields to Nuclear Phenomena},
Eds., J.A. McNeil and C.E. Price (World Scientific, Singapore 1991)
pp. 46-54;
R.L. Jaffe and P.F. Mende, Nucl. Phys. B {\bf 369}, 189 (1992).

\bibitem{cahill92}
R.T. Cahill, Nucl. Phys. A {\bf 543}, 63 (1992).

\bibitem{cahill85}
R.T. Cahill and C.D. Roberts, Phys. Rev. D {\bf 32}, 2419 (1985).

\bibitem{kleinert78}
H. Kleinert, Phys. Lett. B {\bf 62}, 429 (1976); in {\it Proc. of the 1976
School of Subnuclear Physics}, Erice (Plenum, 1978); Fortschr.
Phys. {\bf 30}, 351 (1982).

\bibitem{cahill89}
R.T. Cahill, Aust. J. Phys. {\bf 42}, 171 (1989).

\bibitem{roberts88}
C.D. Roberts, R.T. Cahill, and J. Praschifka, Ann. Phys. (N.Y.)
{\bf 188}, 20 (1988).

\bibitem{roberts94}
C.D. Roberts, R.T. Cahill, M.E. Sevior, and N. Iannella, Phys.
Rev. D {\bf 49}, 125 (1994).

\bibitem{frank94a}
M.R. Frank and P.C. Tandy, Phys. Rev. C {\bf 49}, 478 (1994).

\bibitem{holstein}John F. Donoghue, Eugene Golowich, and Barry R. Holstein,
{\it Dynamics of the Standard Model} (Cambridge University Press, 1992), and
references therein.

\bibitem{hatsuda}T. Hatsuda, Phys. Rev. Lett. {\bf 65}, 543 (1990).


\bibitem{meissner95}T. Meissner, Phys. Lett. B {\bf 340}, 226 (1994).

\bibitem{witten}E. Witten, Nucl. Phys. B{\bf 156}, 269 (1979).

\bibitem{frank95c}M.R. Frank, B.K. Jennings, and G.A. Miller, ``The
role of color neutrality in nuclear physics: Modifications of the
nucleonic wave functions", submitted to Physical Review C (1995), and
references therein.

\bibitem{nambu61}
Y. Nambu and G. Jona-Lasinio, Phys. Rev. {\bf 122}, 345 (1961);
{\bf 124}, 246 (1961).


\bibitem{ruizarriola}
E. Ruiz Arriola, Phys. Lett. B {\bf 253}, 430 (1991);
Phys. Lett. B {\bf 264}, 178 (1991);
C. Schueren, E. Ruiz Arriola and K. Goeke, Nucl. Phys. A {\bf 547}, 612 (1993).

\bibitem{bijnens}
J. Biijnens, C. Bruno and E. de Rafael, Nucl. Phys. B {\bf 390}, 501 (1993);
J. Bijnens, ``Chiral Lagrangians and Nambu-Jona-
Lassino-like models", NORDITA-95-10-N-P, Feb 1995.

\bibitem{klevansky}
J. Mueller and S. Klevansky, Phys. Rev. C {\bf 50}, 410 (1994).

\bibitem{shrauner}E. Shrauner, Phys. Rev. D{\bf 16},1887 (1977).

\bibitem{gsm}
M.R. Frank, P.C. Tandy, and G. Fai, Phys. Rev. C {\bf 43}, 2808 (1991);
P.C. Tandy and M.R. Frank, Aust. J. Phys. {\bf 44}, 181 (1991);
M.R. Frank and P.C. Tandy, Phys. Rev. C {\bf 46}, 338 (1992).

\bibitem{frank95b}M.R. Frank in {\it Few-Body Problems in Physics},
Ed.,Franz Gross ( AIP Conference Proceedings 334)
pp. 15-30.

\bibitem{formfactors}C.D. Roberts, {it Electromagnetic Pion Form
Factor and Neutral Pion Decay Width}, ANL preprint no.
ANL-PHY-7842-TH-94 (1994);  M.R, Frank, K.L. Mitchell,
C.D. Roberts, and P.C. Tandy, Phys. Lett. B {\bf 359}, 17 (1995);
C. J. Burden, C. D. Roberts and M. J. Thomson, {\it
Electromagnetic Form Factors of Charged and Neutral Kaons},
ANL preprint no. ANL-PHY-8240-TH-95 (1995).

\bibitem{hollenberg92alkofer93}
L.C.L. Hollenberg, C.D. Roberts, and B.N.J. McKellar,
Phys. Rev. C {\bf 46}, 2057 (1992);
R. Alkofer, A. Bender, and C.D. Roberts,
``Pion loop contribution to the electromagnetic pion
charge radius", ANL-PHY-7663-TH-93, UNITUE-THEP-13/1993 (1993).

\bibitem{delbourgo79}R. Delbourgo and M.D. Scadron, J. Phys. G
{\bf 5}, 1621 (1979).

\bibitem{review}
C.D. Roberts and A.G. Williams, {\it Dyson-Schwinger equations and their
application to
hadronic physics},  in {\it Progress in Particle and Nuclear Physics}, Ed., A.
Faessler
(Pergamon Press, Oxford 1994); and references therein.


\bibitem{praschifka89}
J. Praschifka, R.T.Cahill, and C.D.Roberts, Int. J. Mod. Phys. A {\bf 4}, 4929
(1989).

\bibitem{frank96a}M.R. Frank and C.D. Roberts,``Model gluon
propagator and pion and rho-meson observables", to appear in Physical
Review C, January (1996).

\bibitem{marciano78}
W. Marciano and H. Pagles, Phys. Rep. C {\bf 36}, 137 (1978).

\bibitem{frank95a}
M.R. Frank, Phys. Rev. C {\bf 51}, 987 (1995).


\end{references}
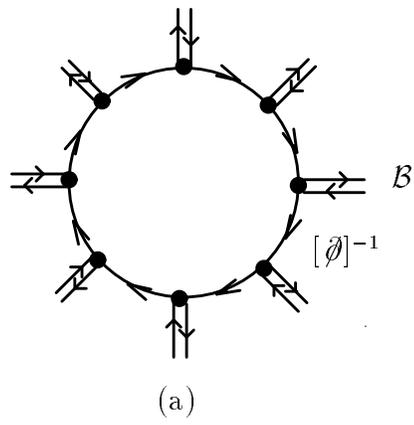
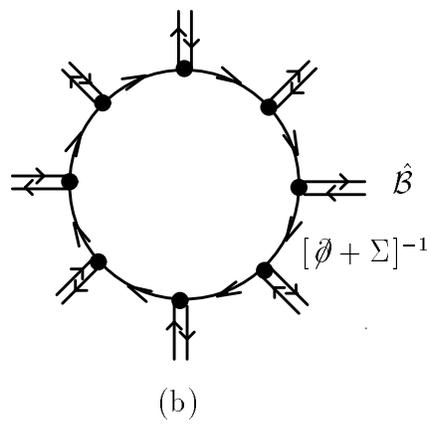
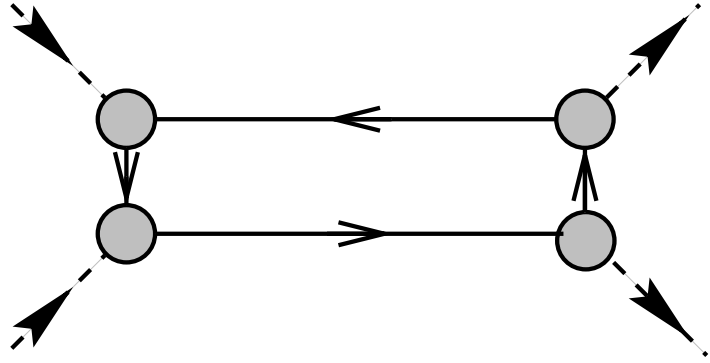
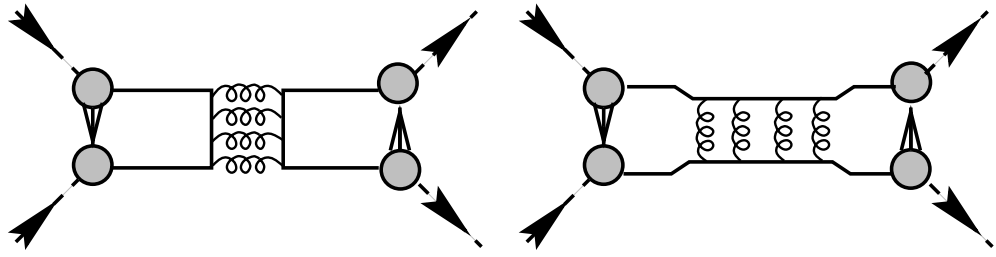
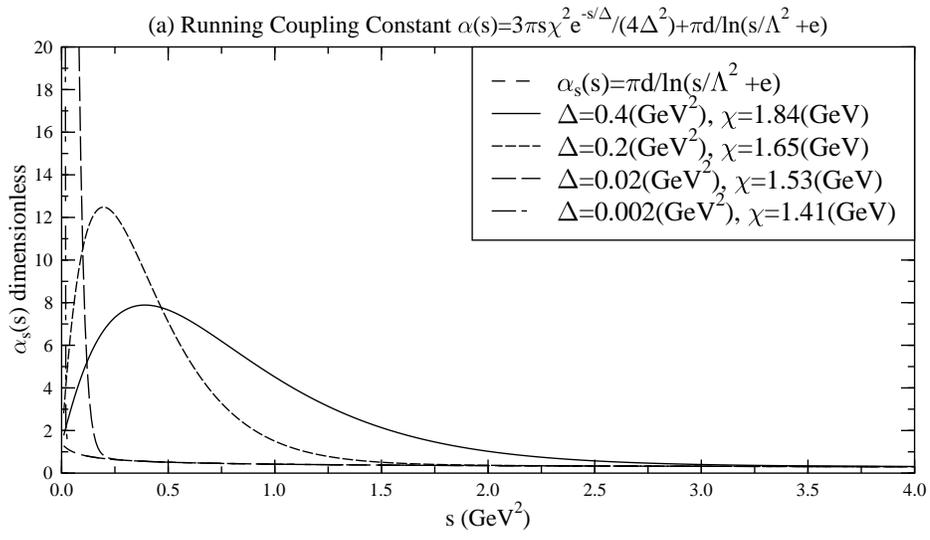
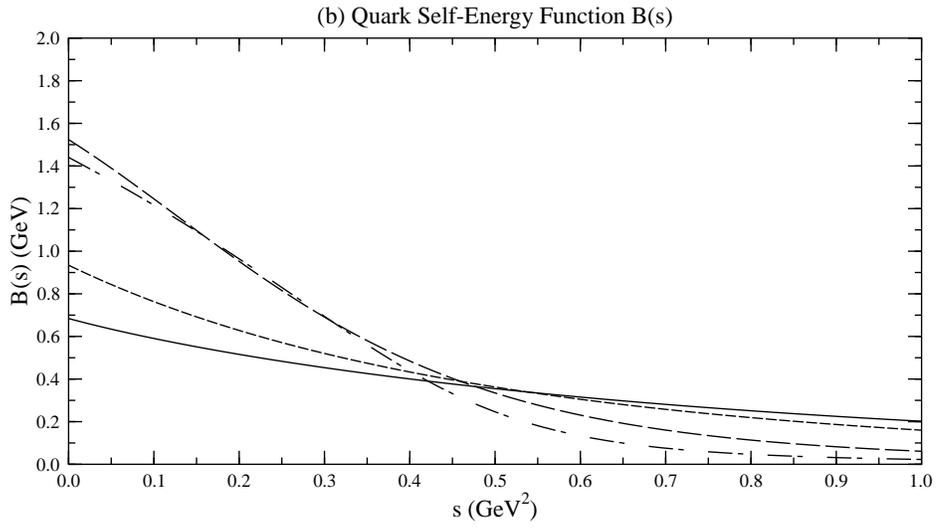
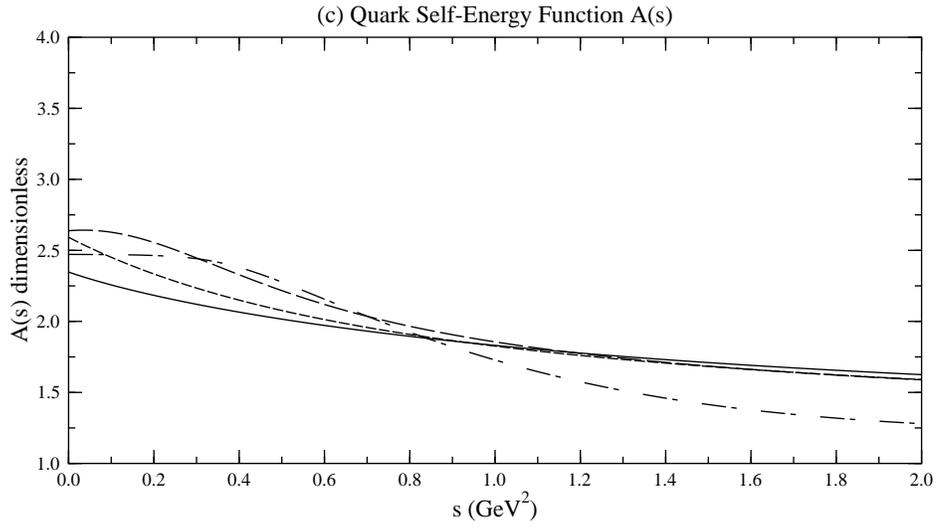
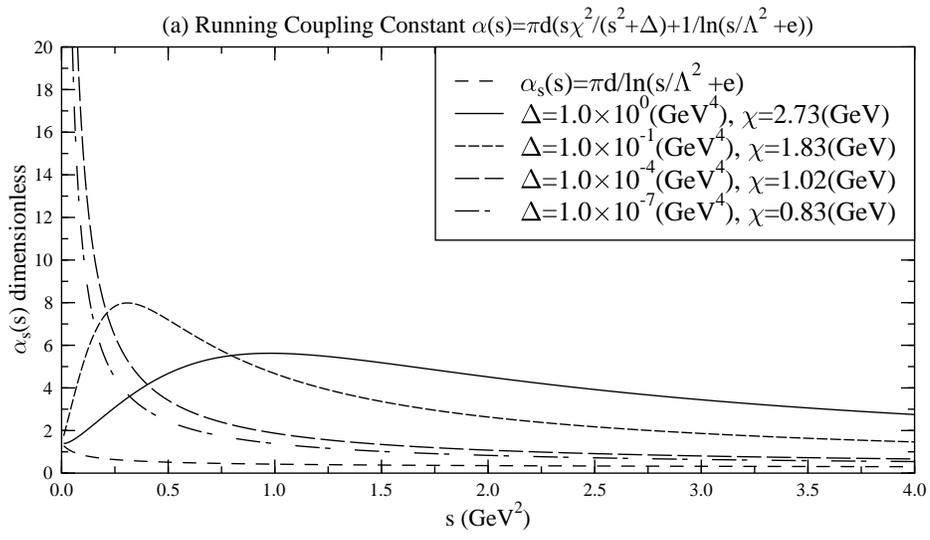
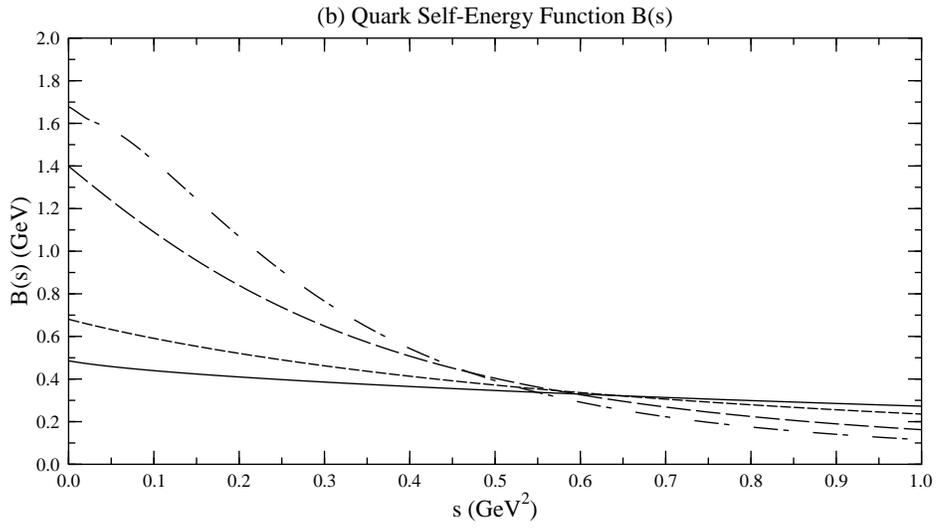
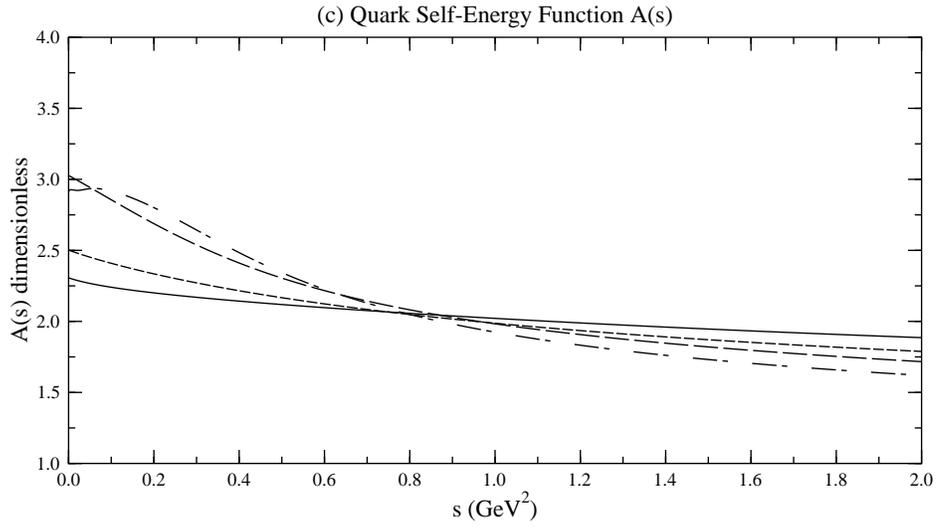
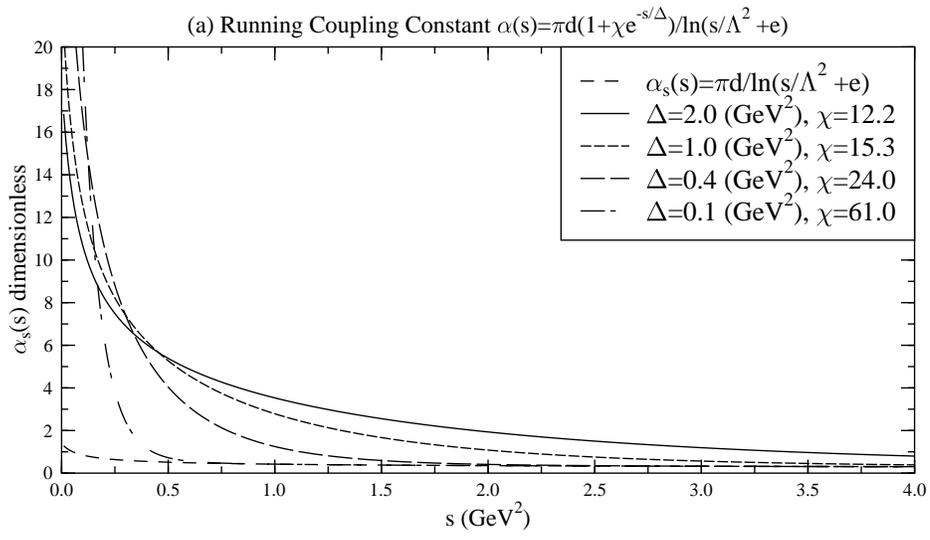
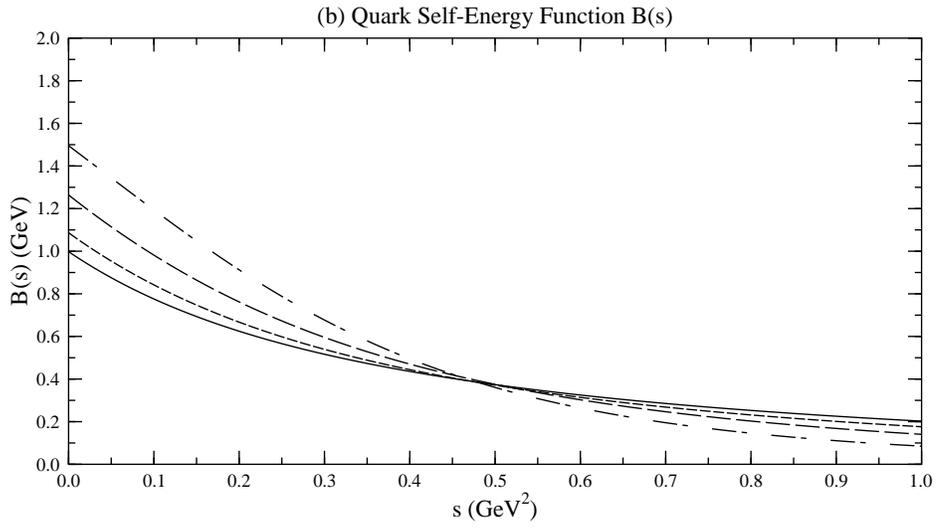
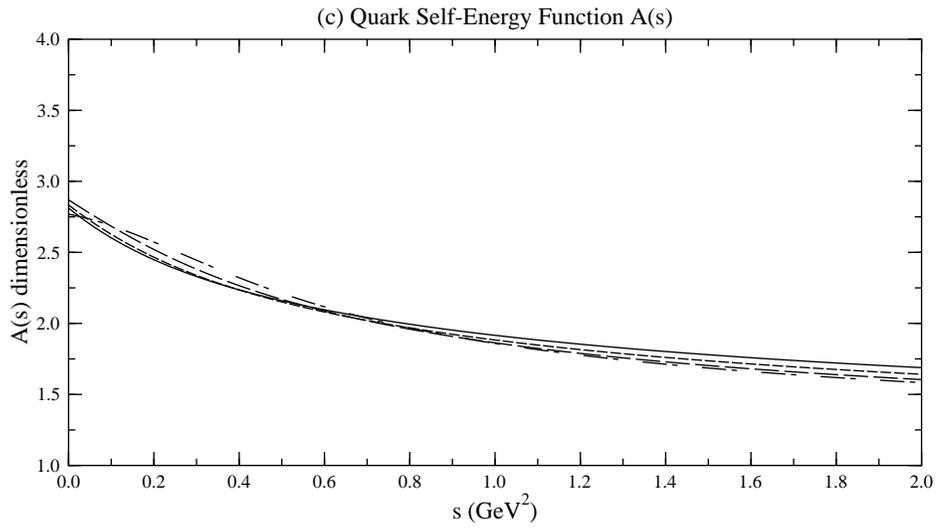
\end{document}